\documentclass[11pt]{article}

\usepackage{hyperref,amsfonts,amsmath,amssymb,bm,a4wide,cite,subfigure,graphicx}

\numberwithin{equation}{section}

\begin{document}

\title{\textbf{Superfluidity in neutrino clusters}}

\author{Maxim Dvornikov\thanks{maxdvo@izmiran.ru}
\\
\small{\ Pushkov Institute of Terrestrial Magnetism, Ionosphere} \\
\small{and Radiowave Propagation (IZMIRAN),} \\
\small{108840 Moscow, Troitsk, Russia; and} \\
\small{\ Bogoliubov Laboratory of Theoretical Physics,} \\
\small{Joint Institute for Nuclear Research,} \\
\small{141980 Dubna, Moscow region, Russia}}

\date{}

\maketitle

\begin{abstract}
We study the formation of a superfluid condensate of neutrinos inside
a neutrino cluster. The attractive interaction between neutrinos is
mediated by a scalar boson which is lighter than a neutrino. We consider
the appearance of neutrino bound states consisting of particles with oppositely
directed spins. The gap equation for such a system is derived. Based on numerical simulations of the neutrino distribution in a cluster, we find the phase transition temperature and the coherence length inside such a cluster for various parameters of the system. The constraints on the parameters of the Yukawa interaction, resulting in the neutrino superfluidity, are derived. We
obtain that the cosmic neutrino background can contribute to the
superfluid condensate inside a neutrino cluster having realistic characteristics. The mechanism of the neutrino cluster cooling in the early universe, based on the plasmons \v{C}erenkov radiation, is proposed.
\end{abstract}


\section{Introduction}

The search of astronomical objects consisting of new particles, e.g.,
forming dark matter, is a problem for modern astrophysics~\cite{Ili23}.
Star like objects composed of bosons, e.g., of axions, which can be
candidates for dark matter, are discussed in Refs.~\cite{BraZha19,Vic21}.
Virialized axions can form inhomogeneous structures called miniclusters~\cite{KolTka94}.

Neutrinos, especially sterile ones, can also contribute to dark matter~\cite{Boy19}.
The formation of spatially localized objects, called neutrino clusters,
consisting of neutrinos was studied in Refs.~\cite{Ste98,SmiXu22}.
Neutrinos are supposed in Refs.~\cite{Ste98,SmiXu22} to be confined
in a cluster by the exchange of a light scalar particle, which is
beyond the standard model. Note that the possibility for neutrinos
to form bound states inside astrophysical objects even in frames of
the standard model was discussed in Ref.~\cite{KieWei97}. The neutrino
trapping inside the rapidly rotating matter of a neutron star was
studied in Refs.~\cite{BalPopStu97,Dvo14}. The latter phenomenon
is also supposed to take place in frames of the standard model. The
neutrino interaction mediated by $Z$ and $W$ bosons with rotating background
fermions was considered in Refs.~\cite{BalPopStu97,Dvo14} in the
forward scattering approximation.

The neutrino interaction through a scalar particle was shown 
in Ref.~\cite{PesSch95} to be attractive. Such interaction was suggested
in Ref.~\cite{Kap04} to lead to the formation of bound states of
neutrinos having opposite chiralities, analogously to Cooper pairs
of electrons in a metal. Assuming that the scalar particle is the Higgs
boson, the neutrino superfluidity was predicted in Ref.~\cite{Kap04}.
The superfluidity of neutrinos was found to take place in a neutron
star only if neutrinos are heavier than the Higgs boson. Thus, such
neutrinos should be beyond the standard model. The possible phenomenological
consequences of the neutrino superfluidity are studied in Ref.~\cite{Aza11}.

Contrary to Ref.~\cite{Kap04}, we can consider the interaction between
usual active standard model neutrinos mediated by a light scalar
boson to examine the superfluidity in a neutrino cluster discussed in
Ref.~\cite{SmiXu22}. The existence of such light bosons was explored,
e.g., in Ref.~\cite{Ber18}. The study of the neutrino superfluidity
in a system with particles having analogous properties was carried
out in Ref.~\cite{Add22}.

The present work is organized as follows. We start in Sec.~\ref{sec:COND}
with the description of the formation of the neutrino condensate.
Then, in Sec.~\ref{sec:GAP}, we derive the gap equation. The application
of our results to the neutrino superfluidity in a cluster is analyzed
in Sec.~\ref{sec:CLUST}. Finally, in Sec.~\ref{sec:CONCL}, we
conclude. The equation driving the distribution of neutrinos in a cluster is rederived in Appendix~\ref{sec:PROPCLUST}. In Appendix~\ref{sec:COOL}, we propose a possible mechanism for the cooling of a neutrino cluster down in case it is formed in the early universe.

\section{Neutrino condensation\label{sec:COND}}

We start with the Lagrangian for the single neutrino $\nu$ interacting
with the scalar field $\phi$,
\begin{equation}\label{eq:Lagrtot}
  \mathcal{L}=\bar{\nu}(\mathrm{i}\gamma^{\mu}\partial_{\mu}-m)\nu+
  \frac{1}{2}(\partial_{\mu}\phi\partial^{\mu}\phi-M^{2}\phi^{2})+g\phi\bar{\nu}\nu,
\end{equation}
where $m$ and $M$ are the masses of a neutrino and a scalar boson, and
$g$ is the coupling constant. We assume that $\nu$ is an electron neutrino for the definiteness.
Writing down the field equation for $\phi$ and assuming that it is
independent of the coordinates, $\partial_{\mu}\phi=0$, we get that
$\phi=\frac{g}{M^{2}}\bar{\nu}\nu.$ Returning this $\phi$ back to
Eq.~(\ref{eq:Lagrtot}), we obtain the nonlinear neutrino Lagrangian
in the form,
\begin{equation}
  \mathcal{L}=\bar{\nu}(\mathrm{i}\gamma^{\mu}\partial_{\mu}-m)\nu+\frac{g^{2}}{2M^{2}}(\bar{\nu}\nu)(\bar{\nu}\nu),
\end{equation}
where we include the additional combinatorial factor $1/2$ in the
interaction term.

Representing the neutrino bispinor as $\nu^{\text{T}}=(\varphi,\chi)$,
where $\varphi$ and $\chi$ are the two component spinors, we get
that
\begin{equation}\label{eq:nubarnu}
  \bar{\nu}\nu=\frac{2m}{\varepsilon+m}\varphi^{\dagger}\varphi,
\end{equation}
where $\varepsilon=\sqrt{p^{2}+m^{2}}$ is the energy of a noninteracting
neutrino. If we neglect antineutrinos, we decompose $\varphi$ into
the spin states
\begin{equation}\label{eq:spindecom}
  \varphi=\varphi_{+}+\varphi_{-},
  \quad
  \varphi_{\pm}=\frac{1}{\sqrt{\mathcal{V}}}
  \sum_{\mathbf{p}}u_{\pm}(\mathbf{p})b_{\pm}(\mathbf{p})e^{-\mathrm{i}\varepsilon t+\mathrm{i}\mathbf{px}},
\end{equation}
where $u_{\pm}(\mathbf{p})$ are the basis spinors, $b_{\pm}(\mathbf{p})$
are the anticommuting annihilation operators, and $\mathcal{V}$ is
the normalization volume. We can take that
\begin{equation}\label{eq:upm}
u_{+}(\mathbf{p})=\left(\begin{array}{c}
1\\
0
\end{array}\right),\quad u_{-}(\mathbf{p})=\left(\begin{array}{c}
0\\
1
\end{array}\right),
\end{equation}
if we choose the $z$-axis for the spin quantization.

The neutrino condensation in frames of the standard model mediated
by the Higgs boson exchange was studied in Ref.~\cite{Kap04}. Chiral
projections of neutrino wavefunctions are inherent in the standard
model. It was the motivation of Ref.~\cite{Kap04} to study the condensation
of the chiral neutrino projections. In our work, we consider the interaction
mediated by the scalar field $\phi$ which is beyond the standard
model. That is why, as in the usual Bardeen--Cooper--Schrieffer
(BCS) superconductivity mechanism (see, e.g., Ref.~\cite{Tin96}),
we suppose that the neutrino condensate is formed by particles with
opposite spins,
\begin{equation}\label{eq:condgap}
  \left\langle \varphi_{-a}(\mathbf{p})\varphi_{+b}(\mathbf{p})\right\rangle =\epsilon_{ab}D.
\end{equation}
where $a,b=1,2$ are the spinor indexes, $D$ is the scalar parameter
to be determined later, and $\epsilon_{ab}$ is the 2D Levi-Civita
symbol.

Using Eqs.~(\ref{eq:nubarnu}) and~(\ref{eq:spindecom}), as well
as applying the Wick theorem, we transform the term in the interaction
Lagrangian to
\begin{align}\label{eq:nubarnu2}
(\bar{\nu}_{\mathbf{p}}\nu_{\mathbf{p}})(\bar{\nu}_{\mathbf{p}}\nu_{\mathbf{p}})\to & \frac{8m^{2}}{\left(\varepsilon+m\right)^{2}}\epsilon_{ab}\nonumber \\
 & \times\left[D^{*}\left(\varphi_{+a}\varphi_{-b}+\varphi_{-a}\varphi_{+b}\right)-D\left(\varphi_{+a}^{\dagger}\varphi_{-b}^{\dagger}+\varphi_{-a}^{\dagger}\varphi_{+b}^{\dagger}\right)\right].
\end{align}
We use Eq.~(\ref{eq:nubarnu2}) in the interaction Hamiltonian $H_{\mathrm{int}}=\int\mathrm{d}^{3}x\mathcal{H}_{\mathrm{int}}$,
where $\mathcal{H}_{\mathrm{int}}=-\mathcal{L}_{\mathrm{int}}$, which
takes the form,
\begin{align}
  H_{\mathrm{int}}= & -\frac{4g^{2}}{M^{2}}\sum_{\mathbf{p}}\frac{m^{2}}{\left(\varepsilon+m\right)^{2}}
  \nonumber
  \\
  & \times
  \Big\{
    D^{*}e^{-2\mathrm{i}\varepsilon t}
    \left[
      b_{+}(\mathbf{p})b_{-}(-\mathbf{p})-b_{-}(\mathbf{p})b_{+}(-\mathbf{p})
    \right]
    \nonumber
    \\
    & +
    De^{2\mathrm{i}\varepsilon t}
    \left[
      b_{-}^{\dagger}(\mathbf{p})b_{+}^{\dagger}(-\mathbf{p})-b_{+}^{\dagger}(\mathbf{p})b_{-}^{\dagger}(-\mathbf{p})
    \right]
  \Big\},
\end{align}
where we use Eqs.~(\ref{eq:spindecom}) and~(\ref{eq:upm}). The total Hamiltonian $H = H_0 + H_{\mathrm{int}}$ should contain the noninteracting term,
\begin{equation}
  H_{0}=\sum_{\mathbf{p}}(\varepsilon-\mu)
  \left[
    b_{+}^{\dagger}(\mathbf{p})b_{+}(\mathbf{p})+b_{-}^{\dagger}(\mathbf{p})b_{-}(\mathbf{p})
  \right],
\end{equation}
where $\mu$ is the chemical potential.  

We suppose that $D=|D|e^{2\mathrm{i}\alpha}$, where $\alpha$ is
the phase. Making the Bogoliubov transformations
\begin{align}
  c_{+}(\mathbf{p}) & =
  e^{-\mathrm{i}\varepsilon t-\mathrm{i}\alpha}\cos\frac{\theta}{2}b_{+}(\mathbf{p})+
  e^{\mathrm{i}\varepsilon t+\mathrm{i}\alpha}\sin\frac{\theta}{2}b_{-}^{\dagger}(-\mathbf{p}),
  \nonumber
  \\
  c_{-}^{\dagger}(-\mathbf{p}) & =
  -e^{-\mathrm{i}\varepsilon t-\mathrm{i}\alpha}\sin\frac{\theta}{2}b_{+}(\mathbf{p})+
  e^{\mathrm{i}\varepsilon t+\mathrm{i}\alpha}\cos\frac{\theta}{2}b_{-}^{\dagger}(-\mathbf{p}),
  \label{eq:Bt1}
  \\
  \intertext{and}
  c_{-}(\mathbf{p}) & =
  e^{-\mathrm{i}\varepsilon t-\mathrm{i}\alpha}\cos\frac{\theta}{2}b_{-}(\mathbf{p})-
  e^{\mathrm{i}\varepsilon t+\mathrm{i}\alpha}\sin\frac{\theta}{2}b_{+}^{\dagger}(-\mathbf{p}),
  \nonumber
  \\
  c_{+}^{\dagger}(-\mathbf{p}) & =
  e^{-\mathrm{i}\varepsilon t-\mathrm{i}\alpha}\sin\frac{\theta}{2}b_{-}(\mathbf{p})+
  e^{\mathrm{i}\varepsilon t+\mathrm{i}\alpha}\cos\frac{\theta}{2}b_{+}^{\dagger}(-\mathbf{p}),
  \label{eq:Bt2}
\end{align}
where $c_{\pm}(\mathbf{p})$ are the new operators, we bring $H$ to the form,
\begin{align}\label{eq:Hc}
  H= & \sum_{\mathbf{p}}
  \Big\{
    \left[
      (\varepsilon-\mu)\cos\theta+\Delta\sin\theta
    \right]
    \left[
      c_{+}^{\dagger}(\mathbf{p})c_{+}(\mathbf{p})+c_{-}^{\dagger}(\mathbf{p})c_{-}(\mathbf{p})
    \right]
    \nonumber
    \\
    & +
    \left[
      \Delta\cos\theta-\sin\theta(\varepsilon-\mu)
    \right]
    \left[
      c_{+}^{\dagger}(\mathbf{p})c_{-}^{\dagger}(-\mathbf{p})+c_{-}(-\mathbf{p})c_{+}(\mathbf{p})
    \right]
    \nonumber
    \\
    & +
    (\varepsilon-\mu)(1-\cos\theta)-\Delta\sin\theta
  \Big\},
\end{align}
where
\begin{equation}\label{eq:gap}
  \Delta=\frac{8g^{2}m^{2}|D|}{M^{2}\left(\varepsilon+m\right)^{2}},
\end{equation}
is the magnitude of the energy gap.

Then, we take that 
\begin{equation}
  \sin\theta=\frac{\Delta}{E},\quad\cos\theta=\frac{(\varepsilon-\mu)}{E},\quad\tan\theta=\frac{\Delta}{\varepsilon-\mu},
\end{equation}
where $E=\sqrt{(\varepsilon-\mu)^{2}+\Delta^{2}}$. In this situation,
the Hamiltonian in Eq.~(\ref{eq:Hc}) takes the canonical form,
\begin{equation}\label{eq:Hcan}
  H = \sum_{\mathbf{p}}E
  \left[
    c_{+}^{\dagger}(\mathbf{p})c_{+}(\mathbf{p})+c_{-}^{\dagger}(\mathbf{p})c_{-}(\mathbf{p})
  \right],
\end{equation}
where one can see that $E$ is the quasiparticle energy. Note that
we discard the part of the Hamiltonian in Eq.~(\ref{eq:Hcan}) which
does not contain operators.

\section{Gap equation\label{sec:GAP}}

To determine the gap in Eq.~(\ref{eq:gap}) we multiply Eq.~(\ref{eq:condgap})
by $\epsilon_{ab}$, average it over the ground state $\left\langle \dots\right\rangle _{\text{GS}}$,
and integrate over space. Eventually, we get that
\begin{align}
\mathcal{V}D= & -\frac{e^{2\mathrm{i}\alpha}}{4}\sum_{\mathbf{p}}\sin\theta\left\langle c_{+}^{\dagger}(\mathbf{p})c_{+}(\mathbf{p})+c_{-}^{\dagger}(\mathbf{p})c_{-}(\mathbf{p})-1\right\rangle _{\text{GS}},
\end{align}
where we use Eqs.~(\ref{eq:spindecom}), (\ref{eq:upm}), (\ref{eq:Bt1}),
and~(\ref{eq:Bt2}). Then, we notice that
\begin{equation}
\left\langle c_{+}^{\dagger}(\mathbf{p})c_{+}(\mathbf{p})+c_{-}^{\dagger}(\mathbf{p})c_{-}(\mathbf{p})\right\rangle _{\text{GS}}=2n_{f},
\end{equation}
where $n_{f}=1/(e^{E/T}+1)$ is the Fermi-Dirac distribution of quasiparticles
and $T$ is the neutrino gas temperature. Here, we account for two
polarizations of quasiparticles. Despite the chemical potential of the neutrino gas is nonzero, we take the distribution of quasiparticles with  zero chemical potential, as mentioned in Ref.~\cite{LifPit81}. It is also convenient to express
$2n_{f}-1=-\tanh(E/2T)$.

Based on the the fact that
\begin{equation}
  \sin\theta=\frac{4g^{2}m^{2}|D|}{M^{2}
  \left(
    \varepsilon+m
  \right)^{2}}
  \frac{1}{\sqrt{(\varepsilon-\mu)^{2}+\Delta^{2}}},
\end{equation}
and replacing the sum over $\mathbf{p}$ with the 3D integration, $\frac{1}{\mathcal{V}}\sum_{\mathbf{p}}=\int\frac{\mathrm{d}^{3}p}{(2\pi)^{3}}$,
we obtain the gap equation,
\begin{equation}\label{eq:gapeq}
  \frac{2g^{2}}{M^{2}}
  \int\frac{\mathrm{d}^{3}p}{(2\pi)^{3}}\frac{m^{2}}{
  \left(
    \varepsilon+m
  \right)^{2}}
  \frac{\tanh(E/2T)}{\sqrt{(\varepsilon-\mu)^{2}+\Delta^{2}}}=1.
\end{equation}
Note that Eq.~(\ref{eq:gapeq}) contains the factor $\left(\varepsilon+m\right)^{2}$
in the denominator of the integrand versus $\varepsilon^{2}$ in Ref.~\cite{Kap04}.
It is the consequence of the condensation of spin states in our case
rather than chiral ones considered in Ref.~\cite{Kap04}. The presence of the term $m^2$ in the numerator of the integrand in Eq.~\eqref{eq:gapeq} confirms the result of Ref.~\cite{Kap04} that the condensation is possible for massive neutrinos only.

We suppose that the neutrino gas is degenerate, $T\ll E$. The difference
between the energy spectra of quasiparticles and unpaired neutrinos
is the most significant near the Fermi surface~\cite{Tin96}. Hence, we modify the integral in Eq.~\eqref{eq:gapeq} as
\begin{equation}\label{eq:gapeqmod}
  \frac{m^{2}p_\mathrm{F}^2}
  {2\pi^2
  \left(
    \mu+m
  \right)^{2}}
  \int_0^\Lambda\frac{\mathrm{d}p}{\sqrt{v_\mathrm{F}^2(p-p_\mathrm{F})^{2}+\Delta^{2}}}=
  \frac{m^{2}p_\mathrm{F}^2}
  {2\pi^2 v_\mathrm{F}
  \left(
    \mu+m
  \right)^{2}}
  \int_{-v_\mathrm{F}p_\mathrm{F}}^{v_\mathrm{F}(\Lambda - p_\mathrm{F})}
  \frac{\mathrm{d}x}{\sqrt{x^2+\Delta^{2}}},
\end{equation}
where $p_{\mathrm{F}}$ is the Fermi momentum and $v_{\mathrm{F}}=\left.\frac{\mathrm{d}\varepsilon}{\mathrm{d}p}\right|_{p=p_\mathrm{F}}=\frac{ p_\mathrm{F} }{\mu}$
is the Fermi velocity. The integral in Eq.~\eqref{eq:gapeq} is divergent at great momenta. That is why, in Eq.~\eqref{eq:gapeqmod}, we restrict the momentum integration by the range $0<|\mathbf{p}|<\Lambda$, where $\Lambda$ is the cut-off parameter. 

In case of the electron superconductivity in metals, the value of the cut-off is $\Lambda \sim \omega_\mathrm{D}$, where $\omega_\mathrm{D}$ is the Debye frequency. The magnitude of $\Lambda$ for the Yukawa interaction, given in Eq.~\eqref{eq:Lagrtot}, was mentioned in Ref.~\cite{GarTutVan22} to be unknown. However, we can use the suggestion of Ref.~\cite{LifPit81} and assume that $\Lambda \sim \mu$, which is the smallest possible cut-off.

Based on Eq.~\eqref{eq:gapeqmod}, we find the solution of Eq.~\eqref{eq:gapeq} in the form,
\begin{align}\label{eq:gapLambda}
  \Delta = & \frac{v_\mathrm{F}}{\sinh y}
  \sqrt{ (\Lambda - p_\mathrm{F})^2 + p_\mathrm{F}^2 + 2p_\mathrm{F}(\Lambda - p_\mathrm{F})\cosh y },
  \notag
  \\
  y = & \frac{\pi^{2}M^{2}(\mu+m)^{2}}{g^{2}m^{2} \mu p_\mathrm{F}}.
\end{align}
If we assume that $y\gg 1$ and $\Lambda \gg p_\mathrm{F}$, as well as consider ultrarelativistic neutrinos, we get in Eq.~\eqref{eq:gapLambda} that $\Delta \approx 2 \Lambda \exp(-y)$~\cite{Tin96}.

The phase transition temperature can be found using the standard technique
(see, e.g., Ref.~\cite{Tin96}) by solving the nonlinear equation,
\begin{equation}
  \int_{0}^{\infty}\mathrm{d}x
  \left[
    \frac{\tanh(x/2)}{x}-\frac{1}{\sqrt{x^{2}+(\Delta/T_{c})^{2}}}
  \right]=0.
\end{equation}
The numerical calculation gives one that $T_{c}\approx0.57\Delta$.

\section{Superfluidity of clustered neutrinos\label{sec:CLUST}}

The formation of clusters of neutrinos owing to the exchange by a
new light scalar boson was studied in Ref.~\cite{SmiXu22}. The size
of clusters can vary in a quite broad range and reach a Mpc. The maximal neutrino density in a cluster
can be $\sim10^{9}\,\text{cm}^{-3}$. Such
density corresponds to the neutrino mass $m=0.1\,\text{eV}$, which
is taken as a reference mass in Ref.~\cite{SmiXu22}. It should be
noted that neutrinos are considered as relativistic particles in the center of a cluster and nonrelativistic ones towards its edge.
Thus, we should account for $p_{\mathrm{F}}$ exactly.

\begin{figure}[htbp]
  \centering
  \subfigure[]
  {\label{fig:f1a}
  \includegraphics[scale=.3]{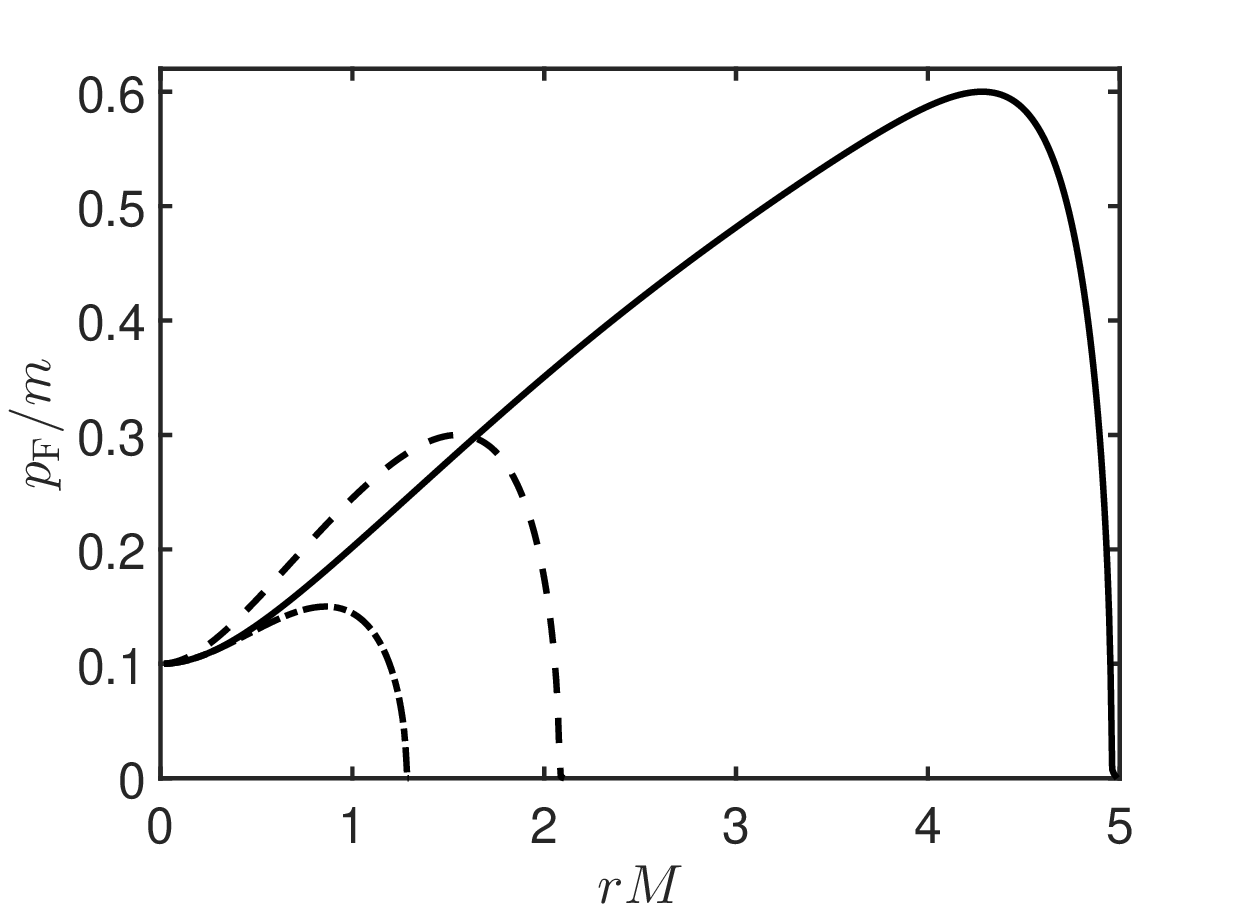}}
  \hskip-.5cm
  \subfigure[]
  {\label{fig:f1b}
  \includegraphics[scale=.3]{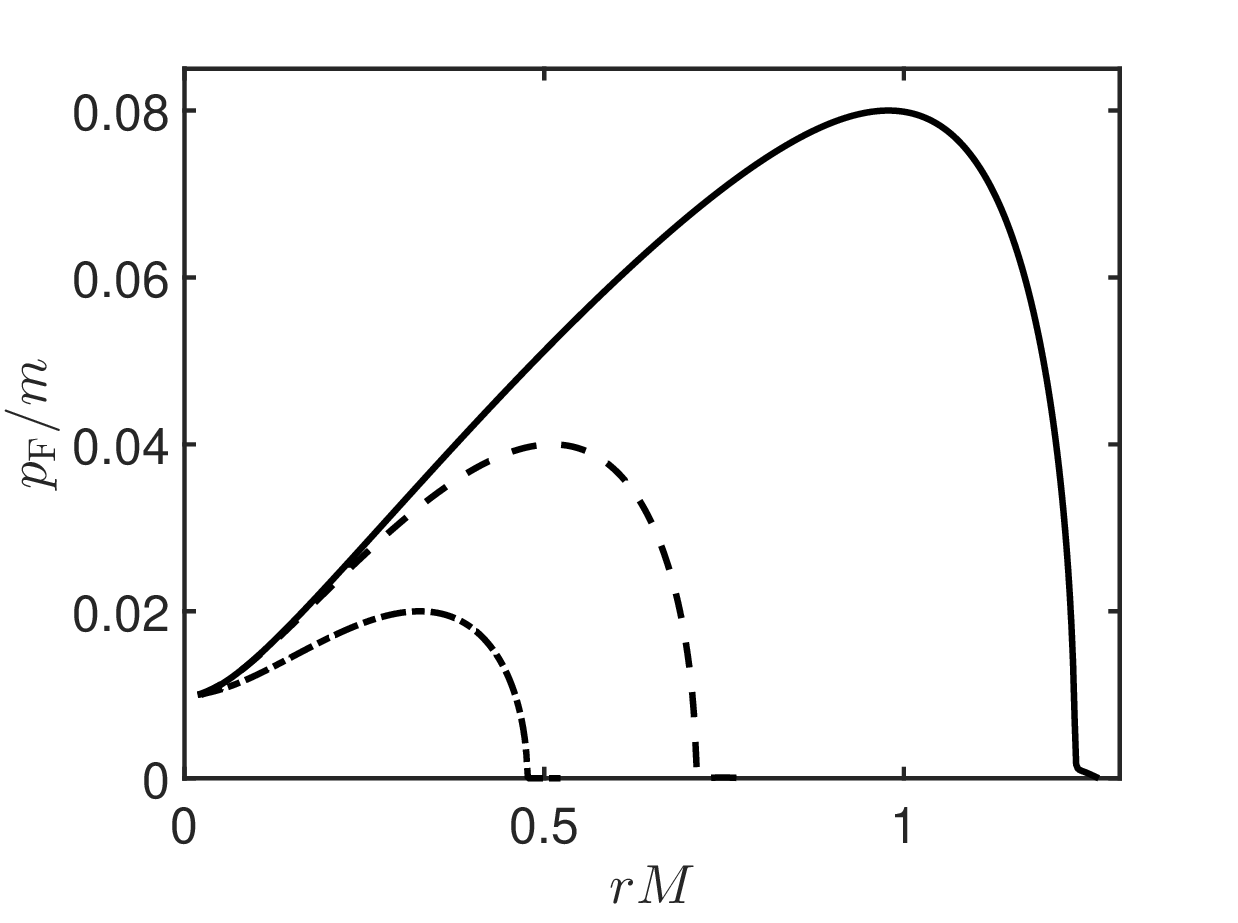}}
  \\
  \subfigure[]
  {\label{fig:f1c}
  \includegraphics[scale=.3]{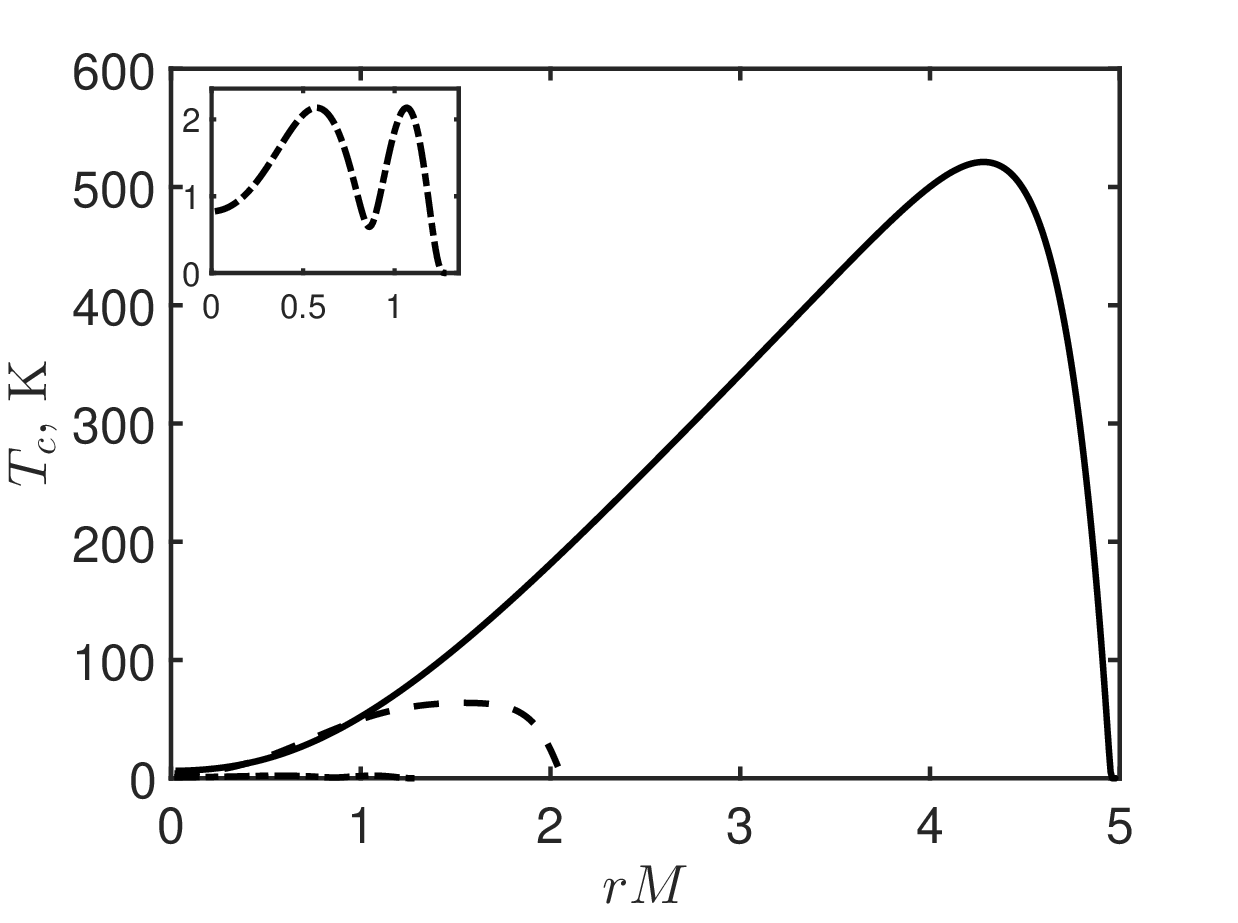}}
  \hskip-.5cm
  \subfigure[]
  {\label{fig:f1d}
    \includegraphics[scale=.3]{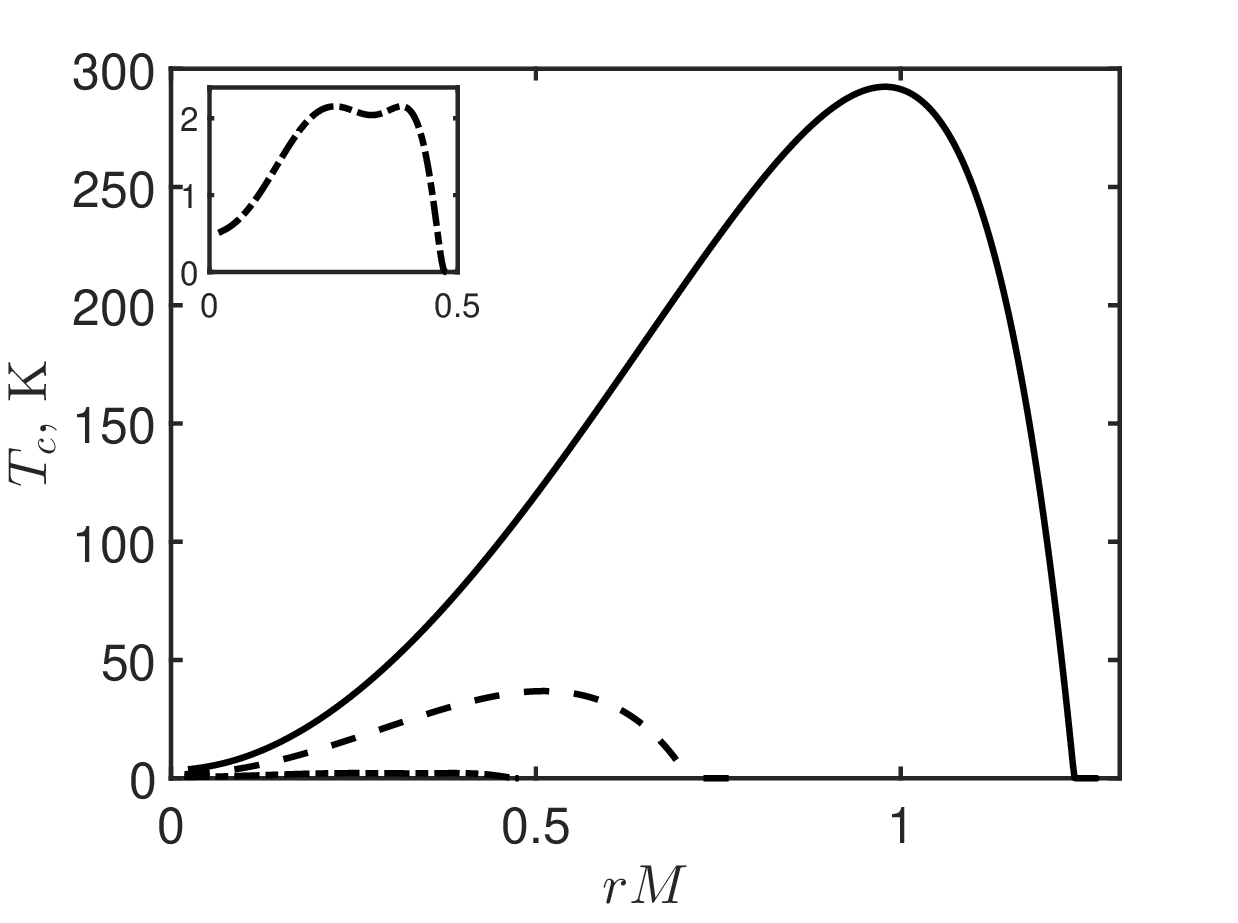}}
   \\
  \subfigure[]
  {\label{fig:f1e}
  \includegraphics[scale=.3]{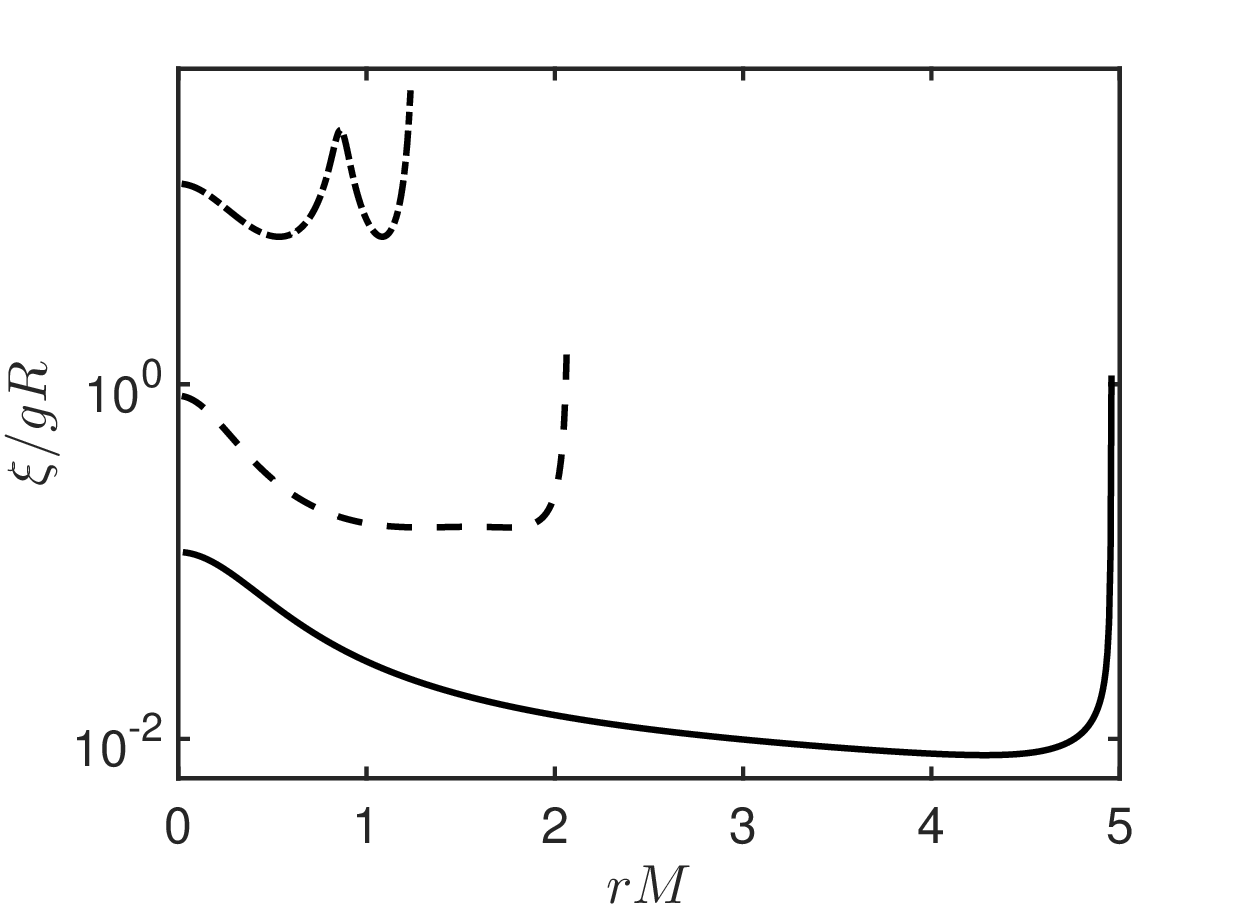}}
  \hskip-.5cm
  \subfigure[]
  {\label{fig:f1f}
  \includegraphics[scale=.3]{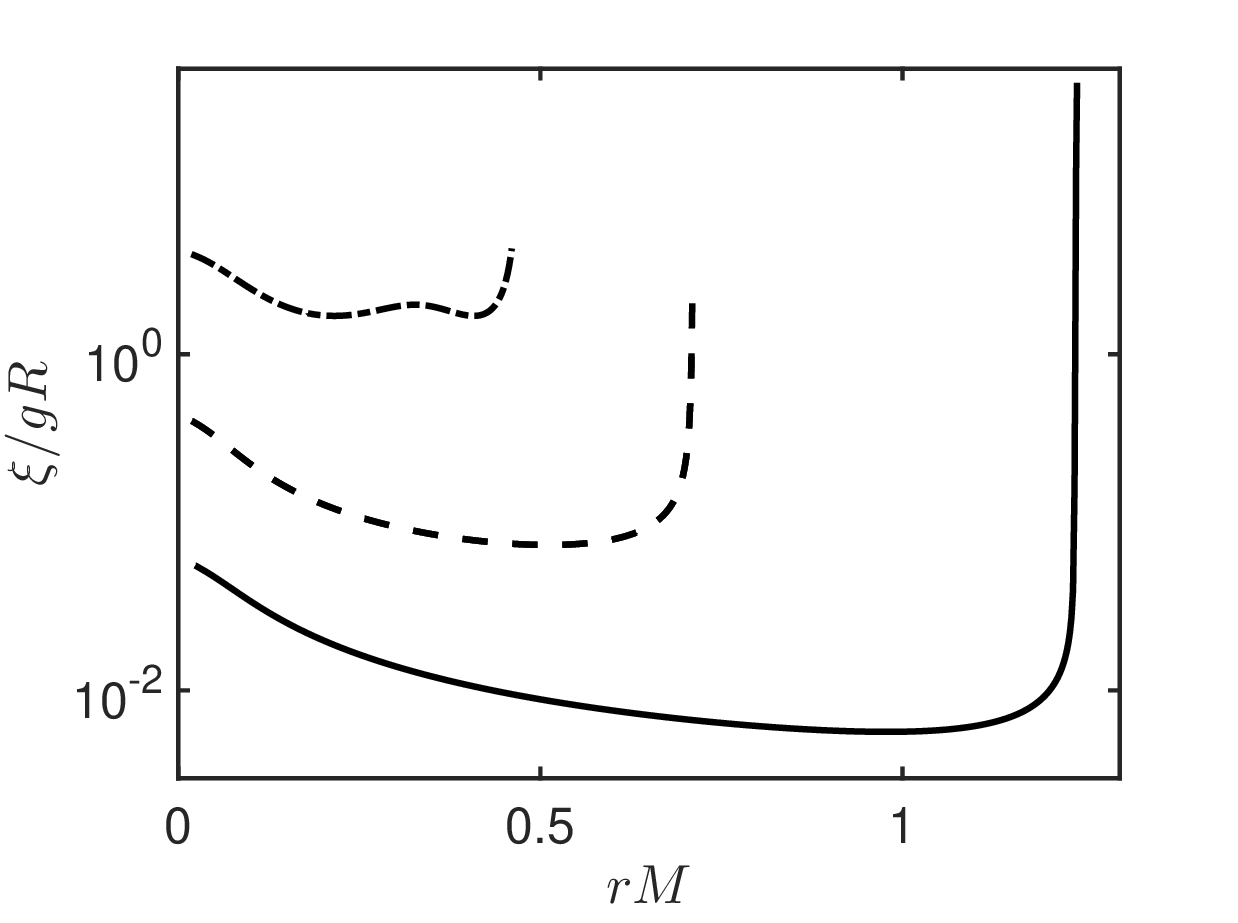}}
  \protect 
\caption{The characteristics of the neutrino gas inside a neutrino cluster as the functions of $r$ for different $\mu_0$, $p_\mathrm{F}(r=0)$, and $\chi$. Panels~(a) and~(b): the normalized Fermi momentum $p_\mathrm{F}(r)/m$; panels~(c) and~(d): the phase transition temperature $T_c(r)$ in Eq.~\eqref{eq:tildeTc}; panels~(e) and~(f): the normalized coherence length $\xi(r)/gR$ in Eq.~\eqref{eq:tildexi}. Panels~(a), (c), and~(e) correspond to $\chi = 0.1$ and $\Pi_0 = 0.1$; solid lines stay for $\mathcal{M} = 0.6$, dashed lines -- for $\mathcal{M} = 0.3$, and dash-dotted ones -- for $\mathcal{M} = 0.15$. Panels~(b), (d), and~(f) correspond to $\chi = 0.01$ and $\Pi_0 = 0.01$; solid lines stay for $\mathcal{M} = 0.08$, dashed lines -- for $\mathcal{M} = 0.04$, and dash-dotted ones -- for $\mathcal{M} = 0.02$. The insets in panels~(c) and~(d) represent $T_c$ in a magnified scale for $\mathcal{M} = 0.15$ and $\mathcal{M} = 0.02$ respectively.\label{fig:clustprop}}
\end{figure}

The distribution of $p_{\mathrm{F}}(r)$ in a cluster was studied in Ref.~\cite{SmiXu22} in details. Nevertheless we rederive the basic equations in Appendix~\ref{sec:PROPCLUST} for the convenience of a reader. The distribution of scalar particles results from Eq.~\eqref{eq:diffeqPhi} which is solved numerically for the given initial condition near the center of a cluster, $\Pi_0 = p_{\mathrm{F}}(r=0)/m$. The solution of Eq.~\eqref{eq:diffeqPhi} also depends on $\mathcal{M} = \mu_0/m$, where $\mu_0$ is the chemical potential in a cluster, which is a constant quantity, as well as on $\chi = \tfrac{M}{gm}$. We vary $\Pi_0$, $\mathcal{M}$, and $\chi$ to get different types of clusters. Our results are shown in Fig.~\ref{fig:clustprop}.

Should $\Phi$ be the solution of Eq.~\eqref{eq:diffeqPhi}, the function $p_{\mathrm{F}}(r)$ is obtained in Eq.~\eqref{eq:pFr}. The normalized Fermi momenta, $p_{\mathrm{F}}(r)/m$, for different parameters of the system are shown in Figs.~\ref{fig:f1a} and~\ref{fig:f1b}. The radius of a cluster $R$ can be obtained from the equation, $p_{\mathrm{F}}(R) = 0$, which is solved numerically. Note that we do not transform the function $p_{\mathrm{F}}(r)$ into the distribution of the neutrino density since the quantities, which we are interested in, e.g., the energy gap in Eq.~\eqref{eq:gapLambda}, depend on $p_{\mathrm{F}}$.


Based on Eq.~\eqref{eq:gapLambda}, we represent the phase transition temperature $T_c = 0.57 \Delta$, which is obtained in Sec.~\ref{sec:COND}, in the form,
\begin{align}\label{eq:tildeTc}
  T_c = & 6.61\times 10^2\,\text{K} \frac{\Pi}{\mathcal{M}\sinh y}
  \sqrt{ (\mathcal{M} - \Pi)^2 + \Pi^2 + 2\Pi(\mathcal{M} - \Pi)\cosh y },
  \notag
  \\
  y = & \pi^2 \chi^2 \frac{(1+\mathcal{M})^2}{\mathcal{M} \Pi},
\end{align}
where $\mathcal{M}$ is defined earlier and $\Pi$ is given in Eq.~\eqref{eq:pFr}. We show $T_c$ in the neutrino clusters for different parameters of the system in Figs.~\ref{fig:f1c} and~\ref{fig:f1d}, which correspond to $\Pi$ in Figs.~\ref{fig:f1a} and~\ref{fig:f1b}.

Suppose that a neutrino cluster is formed by relic neutrinos with $T = T_{\text{C}\nu\text{B}}$,
where $T_{\text{C}\nu\text{B}}=1.95\,\text{K}$ is the temperature
of the cosmic neutrino background~\cite[p.~135]{GorRub11}. One can see in Figs.~\ref{fig:f1c} and~\ref{fig:f1d} that $T < T_c$ in a spherical layer with $R_\mathrm{in} < r < R_\mathrm{out}$. The neutrino condensation takes place in this layer. However, $R_\mathrm{in} \sim 0$ and $R_\mathrm{out} \sim R$ for almost all parameters in Figs.~\ref{fig:f1c} and~\ref{fig:f1d} except the cases $\mathcal{M} = 0.15$ and $\mathcal{M} = 0.02$, which are shown in the insets. It is interesting to notice that the neutrino condensation is suppressed in the very center of a cluster where $p_{\mathrm{F}}$ is typically smaller; cf. Figs.~\ref{fig:f1a} and~\ref{fig:f1b}.

When a neutrino cluster is formed, the neutrino density increases owing to the attractive interaction mediated by the scalar field. In this situation, the temperature of the neutrino gas also increases. This process can destroy the superfluidity of the neutrino gas if its temperature exceeds $T_c$ given in Eq.~\eqref{eq:tildeTc}. Hence a cooling mechanism is required. The cooling mechanism based on the $\phi$ bremsstrahlung proposed in Ref.~\cite{SmiXu22} gives the cooling time longer than the universe life time. In Appendix~\ref{sec:COOL}, we suggest a possible channel for the cooling a neutrino cluster down based on the emission of plasmons. If a cluster is formed in the early universe in the epoch $20\,\text{GeV} < T < 100\,\text{GeV}$, it provides the efficient cluster cooling.

The important quantity characterizing a superfluid state is the coherence
length~\cite{Tin96}, $\xi=\frac{v_{\mathrm{F}}}{\pi\Delta}$. It is the geometrical
size of a Cooper pair of fermions. It is clear that $\xi$ should
be smaller than the length scale of a system. Thus, we normalize it by the radius of a cluster $R$ and require that $\xi<R$. We represent $\xi$ as
\begin{equation}\label{eq:tildexi}
  \frac{\xi}{gR} = \frac{\chi \sinh y}{\pi z_\mathrm{R} \sqrt{ (\mathcal{M} - \Pi)^2 + \Pi^2 + 2\Pi(\mathcal{M} - \Pi)\cosh y }},
\end{equation}
where $z_\mathrm{R} = MR$ is the dimensionless cluster radius.

We show the normalized coherence length, $\xi/gR$, in Figs.~\ref{fig:f1e} and~\ref{fig:f1f} for different parameters of a cluster, which correspond to Figs.~\ref{fig:f1a} and~\ref{fig:f1b}. One can see in Figs.~\ref{fig:f1e} and~\ref{fig:f1f} that $\xi$ grows outside the superfluid layer, i.e. in the center of a cluster and towards its edge. In these regions, the size of neutrino pairs becomes significantly greater than $R$ leading to the destruction of the superfluid phase.

In our analysis, we made the assumption about the neutrino gas degeneracy. The minimal dimensionless chemical potential in Fig.~\ref{fig:clustprop} is $\mathcal{M} = 0.02$. If $m = 0.1\,\text{eV}$, one gets that $\mu_0 = 2\times 10^{-3}\,\text{eV} = 100\,\text{K}$. If we consider relic neutrinos with $T=1.95\,\text{K}$, the condition $\min(\mu_0) \gg T$ is satisfied, i.e. the neutrino gas is indeed degenerate.

One can see in Figs.~\ref{fig:f1c} and~\ref{fig:f1d} that $T_c$ is high for great $p_\mathrm{F}$. It means that the neutrino superfluidity is more achievable in clusters where the neutrino gas is denser. It happens for great $\chi$. However, we cannot consider very big values of $\chi$ since the cooling time for such clusters would be quite long. The maximal neutrino density $\sim 10^9\,\text{cm}^{-3}$ corresponds to the solid line Fig.~\ref{fig:f1a} (see also Ref.~\cite{SmiXu22}). If we discuss $\chi > 0.1$, there is no possibility for such a cluster to cool down in the early universe, at least, owing to the mechanism suggested in Appendix~\ref{sec:COOL}.

Now, let us study the lower bound for $\chi$ for a cluster where a superfluidity is possible. The numerical solution of Eq.~\eqref{eq:diffeqPhi} exists if $\chi = 10^{-5}$ and $\Pi_0 = 10^{-5}$. The solution is unstable for smaller values of $\chi$. It is owing to the factor $\chi^{-2}$ in the left hand side of Eq.~\eqref{eq:diffeqPhi}. Thus, for small values of $\chi$, this equation becomes very stiff. If we take $\mathcal{M} = 1.5\times 10^{-4}$, the maximal $T_c \approx 2\,\text{K}$. However, $\mu = \mathcal{M} m < T_{\text{C}\nu\text{B}}$ and, hence, the assumption of the degeneracy of the neutrino gas is invalid.

\begin{figure}[htbp]
  \centering
  \subfigure[]
  {\label{fig:f2a}
  \includegraphics[scale=.3]{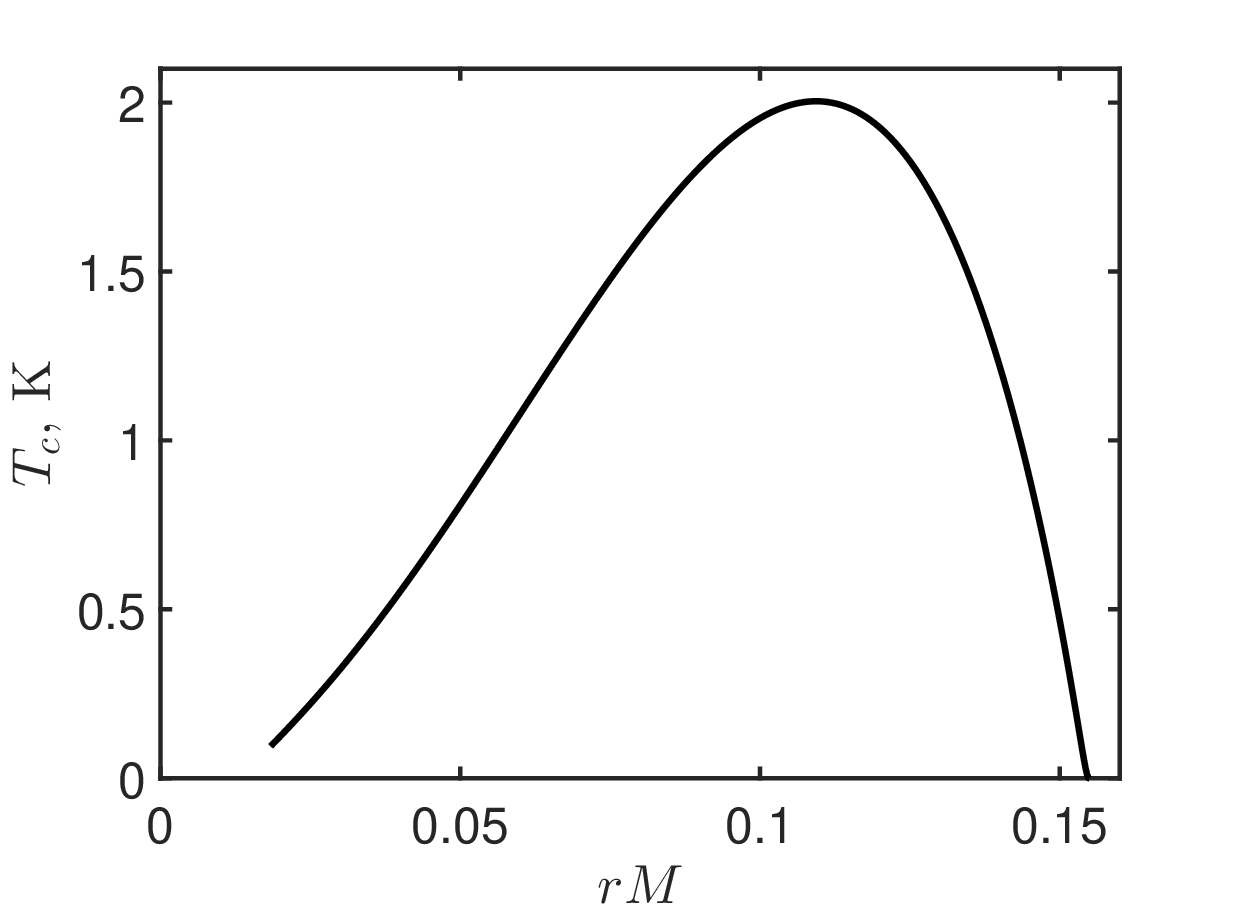}}
  \hskip-.5cm
  \subfigure[]
  {\label{fig:f2b}
  \includegraphics[scale=.3]{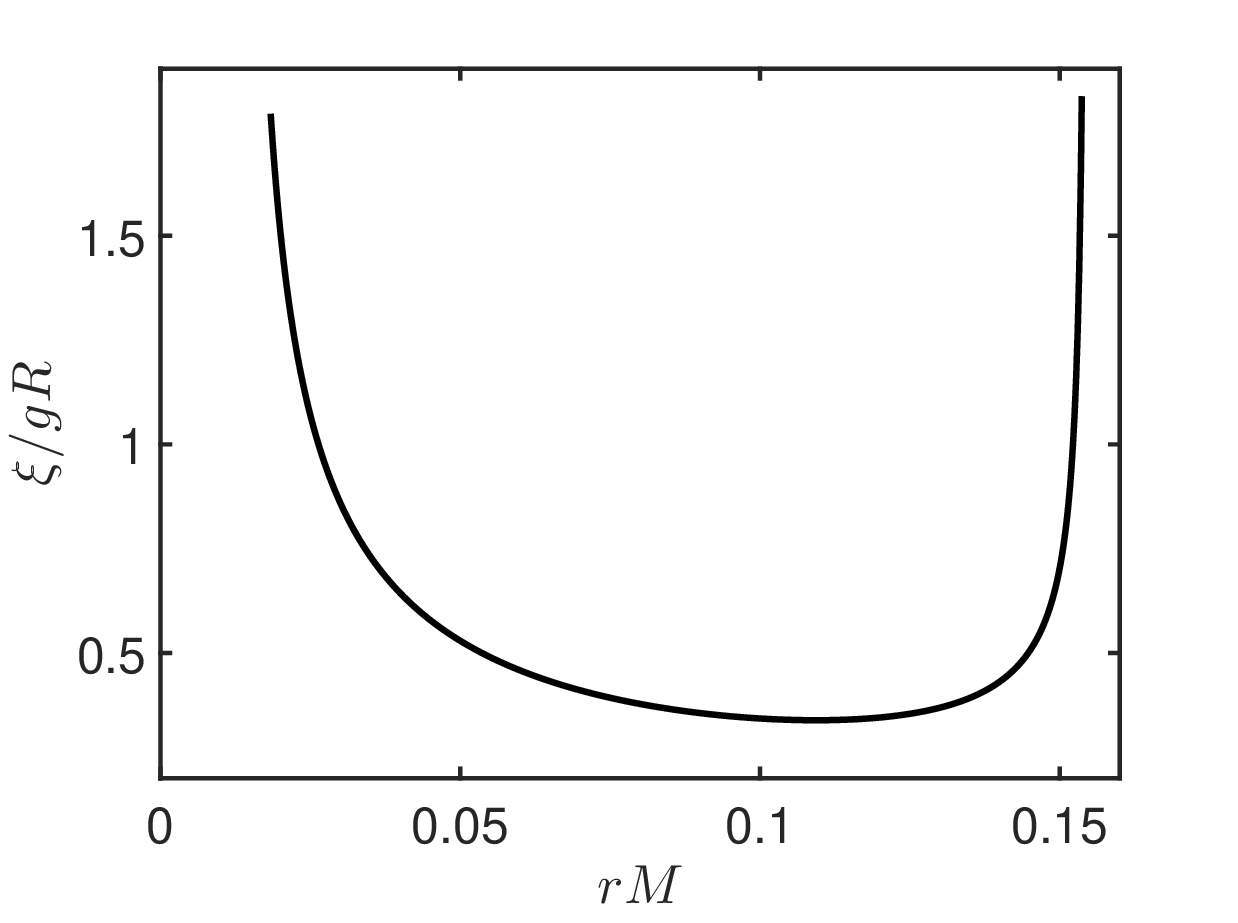}}
  \protect 
\caption{(a) The phase transition temperature and (b) the coherence length for a neutrino cluster with the critical value of $\chi$, where the superfluidity of degenerate neutrinos is still possible. The parameters of the system are $\chi = 5\times 10^{-4}$, $\Pi_0 = 5\times 10^{-4}$, and $\mathcal{M} = 2\times 10^{-3}$ ($\mu = 2.32\,\text{K} > T_{\text{C}\nu\text{B}}$).\label{fig:clustcrit}}
\end{figure}

Taking $\chi = 5\times 10^{-4}$, $\Pi_0 = 5\times 10^{-4}$, and $\mathcal{M} = 2\times 10^{-3}$, we get that the neutrino gas is still superfluid and the condition of the degeneracy is not violated. The phase transition temperature and the coherence length are shown in Fig.~\ref{fig:clustcrit} in this case. Thus, we conclude that the superfluidity in a neutrino cluster is possible when $5\times 10^{-4} < \chi \lesssim 0.1$, provided that this cluster consists of relic neutrinos.

\begin{figure}
  \centering
  \includegraphics[scale=.3]{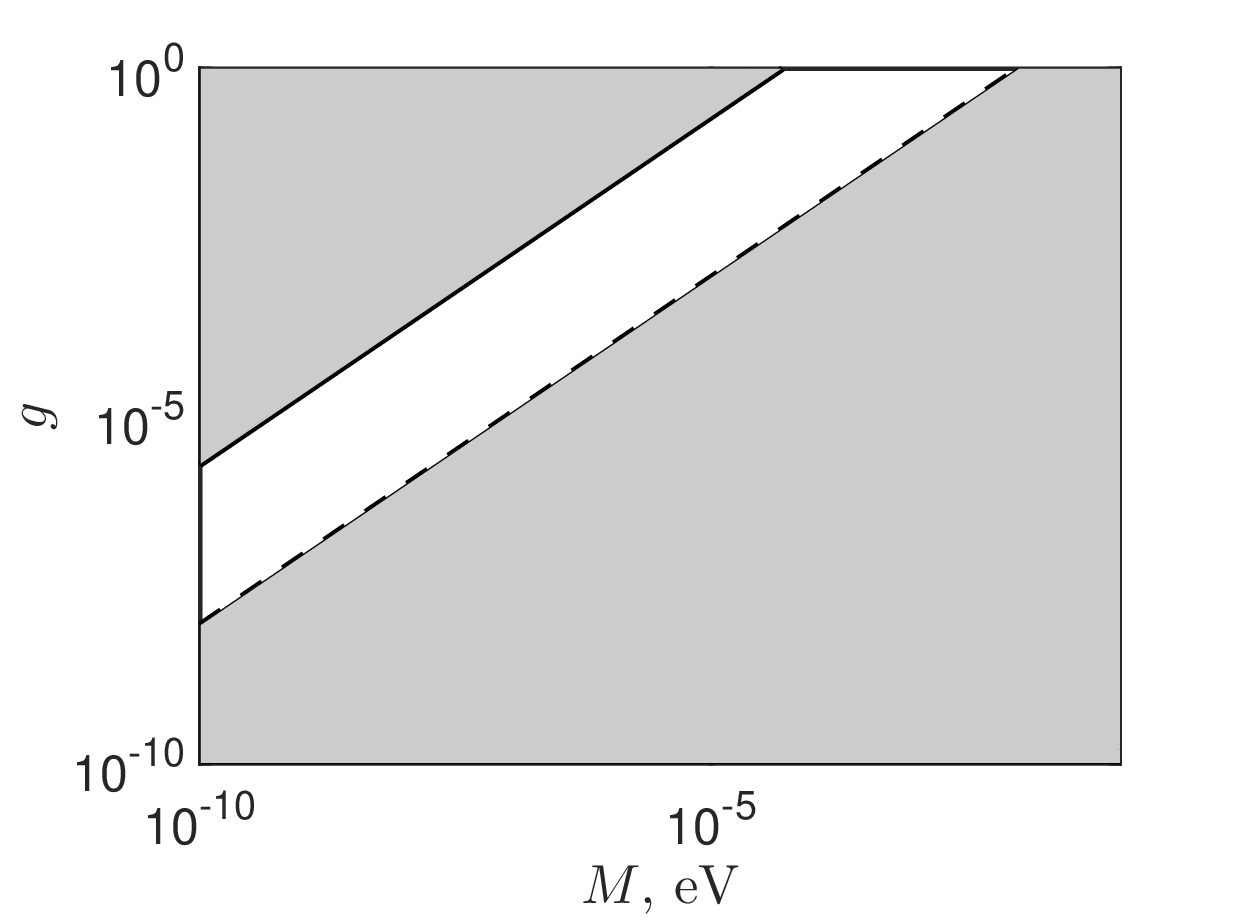}
  \protect 
\caption{The constraints on the parameters $g$ and $M$ leading to the neutrino cluster superfluidity. The shaded areas correspond to the excluded values.\label{fig:gMlim}}
\end{figure}

We can transform the constraints on $\chi$ to the limits on $g$ and $M$. Taking that $m = 0.1\,\text{eV}$, one can finds that $10^2 \left( \tfrac{M}{\text{eV}} \right) < g < 2\times 10^4 \left( \tfrac{M}{\text{eV}} \right)$. The allowed region, where the superfluidity is possible, is depicted in the $(g,M)$-plane in Fig.~\ref{fig:gMlim}. The shaded areas correspond to the excluded values of $g$ and $M$. The lower border of the allowed area is drawn by the dashed line since it depends on a mechanism for the cluster cooling.

The requirement that $\xi < R$, discussed earlier, is fulfilled for any reasonable values of the parameters considered in Figs.~\ref{fig:f1e} and~\ref{fig:f1f}, except the case in Fig.~\ref{fig:f1e} shown by the dash-dotted line. However, if we take the coupling constant in the allowed region in Fig.~\ref{fig:gMlim} (e.g., $g \sim 10^{-5}$), we obtain that $\xi < R$ in this case as well.

Therefore, if a neutrino cluster is formed by cosmic 
neutrinos, e.g., relic ones, and there is an interaction between neutrinos mediated by
a light scalar boson, the neutrino pairing and superfluidity are quite
possible in such a system.

\section{Discussion\label{sec:CONCL}}



In the present work, we have examined the possibility of the superfluidity
inside a neutrino cluster. We have considered a single neutrino eigenstate,
e.g., an electron neutrino with the mass $m=0.1\,\text{eV}$, which
is below the experimental upper bound recently established in Ref.~\cite{Ake22}.
The attractive interaction between neutrinos is mediated by a light
scalar boson. The parameters of the Yukawa interaction, $g$
and $M$, used in our work, are not ruled out by the experimental constraints in these characteristics~\cite{Ber18}.

We have started in Sec.~\ref{sec:COND} with the description of the
neutrino condensation. Contrary to Ref.~\cite{Kap04}, we have assumed
in Eq.~(\ref{eq:condgap}) that the neutrino condensate is formed
by particles with oppositely directed spins like in the usual BCS
mechanism~\cite{Tin96}. Then, in Sec.~\ref{sec:GAP}, we have derived
Eq.~(\ref{eq:gapeq}) for the energy gap which is slightly different
from that in Ref.~\cite{Kap04}.

The application of the obtained results for the neutrino superfluidity
in a cluster, described in Ref.~\cite{SmiXu22}, has been developed
in Sec.~\ref{sec:CLUST}. Based on numerical simulations, present in Appendix~\ref{sec:PROPCLUST}, we have obtained the phase transition temperature and the coherence length inside a neutrino cluster; cf. Figs.~\ref{fig:f1c}-\ref{fig:f1f}. Note that our simulations are valid for both relativistic and nonrelativistic particles. 

Assuming that a neutrino cluster is formed with relic neutrinos, one can see in Figs.~\ref{fig:f1c}-\ref{fig:f1f} that the neutrino superfluidity is quite possible for almost all parameters considered except, maybe, the cases corresponding to insets in Figs.~\ref{fig:f1c} and~\ref{fig:f1d}. We also mention that the superfluid condensate occupies almost all the cluster volume except the very center and the thin layer near the cluster edge.

Since the nonsuperfluid layer surrounding the superfluid core is thin, neutrino pairs still can penetrate it since the condensed phase is not fully destroyed owing to the analogue of the proximity effect (see, e.g., Ref.~\cite[p.~197]{{Tin96}}). It means that superfluid neutrino gas can gradually leak from the neutrino cluster leading to its disappearance. However, this issue requires a separate study.

The constraints on the parameters of the Yukawa interaction between neutrinos and the scalar boson which result in the neutrino superfluidity have been established in Sec.~\ref{sec:CLUST}. One can see in Fig.~\ref{fig:gMlim} that the allowed values of $g$ and $M$ are beyond the sensitivity of current experiments; cf. Ref.~\cite{Ber18}. Therefore, superfluid neutrino clusters are not ruled out. The lower border of the allowed area in Fig.~\ref{fig:gMlim} depends on the cooling mechanism of a cluster, which has been proposed in Appendix~\ref{sec:COOL}.

Finally, we conclude that the superfluidity of cosmic neutrinos
is quite possible in a neutrino cluster. The situation when the Earth appears to be inside
such a cluster is quite probable, accounting for great
sizes of some clusters predicted in Ref.~\cite{SmiXu22}. In this case, one can try to check the existence
of a superfluid condensate in a future $\beta$-decay experiment~\cite{Aza11}.

\appendix

\section{Properties of a neutrino cluster\label{sec:PROPCLUST}}

In this appendix, we review the major steps for the description of
the neutrino cluster properties. Despite we mainly follow the derivation
in Ref.~\cite{SmiXu22}, we represent these results for the convenience of a reader.

The spatial distribution of the neutrino thermodynamic characteristics,
like $p_{\mathrm{F}}$, in case when particles interact with a scalar
boson, can be obtained analogously to $\sigma-\omega$ model in nuclear
physics~\cite{Wal74} if we neglect the contribution of the vector
$\omega$-meson. The wave equations for neutrinos and scalar particles,
resulting from Eq.~\eqref{eq:Lagrtot}, have the form~\cite{Gle00},
\begin{align}
  (\partial_{\mu}\partial^{\mu}+M^{2})\phi & =g\bar{\nu}\nu,
  \label{eq:scalarsigma}
  \\
  (\mathrm{i}\gamma^{\mu}\partial_{\mu}-m+g\phi)\nu & =0.
  \label{eq:Diracsigma}
\end{align}
Using the relativistic mean-field approximation, we get from Eq.~(\ref{eq:Diracsigma})
that the source terms are related by the identity~\cite{Gle00},
\begin{equation}\label{eq:source}
  M^{2}
  \left\langle
    \phi
  \right\rangle =g
  \left\langle
    \bar{\nu}\nu
  \right\rangle.
\end{equation}
Equation~(\ref{eq:source}) means that a neutrino acquires the effective
mass $m_{\mathrm{eff}}=m-g\left\langle \phi\right\rangle $. We can
rewrite Eq.~(\ref{eq:Diracsigma}) in the Hamilton form,
\begin{equation}\label{eq:Heff}
  \varepsilon_{\mathrm{eff}}\nu=H_{\mathrm{eff}}\nu,
  \quad
  H_{\mathrm{eff}}=\bm{\alpha}\mathbf{p}+\beta m_{\mathrm{eff}},
\end{equation}
where $\bm{\alpha}=\gamma^{0}\bm{\gamma}$, $\beta=\gamma^{0}$ are
the Dirac matrices, and
\begin{equation}\label{eq:Eeff}
  \varepsilon_{\mathrm{eff}}=\sqrt{p^{2}+m_{\mathrm{eff}}^{2}}.
\end{equation}
is the effective energy.

We suppose that the neutrino gas is degenerate. Thus, the statistical mean value
of an operator is calculated as 
\begin{equation}\label{eq:meanval}
  \left\langle
    \hat{F}
  \right\rangle =
  2\int_{0}^{p_{\mathrm{F}}}\frac{\mathrm{d}^{3}p}{(2\pi)^{2}}F(\mathbf{p}),
\end{equation}
where $F(\mathbf{p})$ is the eigenvalue of the operator $\hat{F}$.
Note that we take into account the spin degeneracy factor $g_{s}=2$
in Eq.~(\ref{eq:meanval}) despite neutrinos are relativistic particles.
Neutrinos interact by the exchange of a scalar boson. This interaction
mixes the helicity states. Thus, a significant fraction of right neutrinos
appears in the system even if initially we had only left neutrinos.

Then, we use the quantum mechanics theorem for the differentiation
of the Hamiltonian $H$ with respect to any parameter $\varsigma$ (see,
e.g., Ref.~\cite{LanLif65})
\begin{equation}\label{eq:difftheor}
  \left\langle
    n
    \left|
      \frac{\partial H}{\partial\varsigma}
    \right|
    n
  \right\rangle =
  \frac{\partial\varepsilon_{n}}{\partial\varsigma},
\end{equation}
where $\varepsilon_{n}$ is the energy of the state $\left|n\right\rangle $.
Applying Eq.~(\ref{eq:difftheor}) for $\varsigma=m$ and $H=H_{\mathrm{eff}}$
in Eq.~(\ref{eq:Heff}), we get that $\tfrac{\partial H_{\mathrm{eff}}}{\partial m}=\gamma^{0}$
and $\left\langle \bar{\nu}\nu\right\rangle =\left\langle \nu^{\dagger}\left|\gamma^{0}\right|\nu\right\rangle =\tfrac{\partial\varepsilon_{\mathrm{eff}}}{\partial m}$.
Finally, using Eqs.~(\ref{eq:Eeff}) and~(\ref{eq:meanval}), we
obtain that~\cite{Gle00}
\begin{equation}\label{eq:barnunu}
  \left\langle
    \bar{\nu}\nu
  \right\rangle =
  2\int_{0}^{p_{\mathrm{F}}}\frac{\mathrm{d}^{3}p}{(2\pi)^{2}}\frac{\partial}{\partial m}\sqrt{p^{2}+m_{\mathrm{eff}}^{2}}=
  \frac{1}{\pi^{2}}\int_{0}^{p_{\mathrm{F}}}p^{2}\mathrm{d}p\frac{m_{\mathrm{eff}}}{\varepsilon_{\mathrm{eff}}}.
\end{equation}
The integral in Eq.~(\ref{eq:barnunu}) can be computed analytically
giving one
\begin{equation}\label{eq:barnunufin}
  \left\langle
    \bar{\nu}\nu
  \right\rangle =
  \frac{m_{\mathrm{eff}}}{2\pi^{2}}
  \left[
    p_{\mathrm{F}}\sqrt{p_{\mathrm{F}}^{2}+m_{\mathrm{eff}}^{2}}-m_{\mathrm{eff}}^{2}\ln
    \left(
      \frac{p_{\mathrm{F}}+\sqrt{p_{\mathrm{F}}^{2}+m_{\mathrm{eff}}^{2}}}{m_{\mathrm{eff}}}
    \right)
  \right].
\end{equation}
Equation~(\ref{eq:barnunufin}) coincides with the result of Ref.~\cite{SmiXu22}.

We substitute Eq.~(\ref{eq:barnunufin}) to (\ref{eq:scalarsigma})
to determine $\left\langle \phi\right\rangle $ which is supposed to
be time independent,
\begin{multline}\label{eq:diffeqphi}
  \left(
    M^{2}-\frac{\mathrm{d}^{2}}{\mathrm{d}r^{2}}-\frac{2}{r}\frac{\mathrm{d}}{\mathrm{d}r}
  \right)
  \left\langle
    \phi
  \right\rangle =
  \frac{gm_{\mathrm{eff}}}{2\pi^{2}}
  \\
  \times
  \left[
    p_{\mathrm{F}}\sqrt{p_{\mathrm{F}}^{2}+m_{\mathrm{eff}}^{2}}-m_{\mathrm{eff}}^{2}\ln
    \left(
      \frac{p_{\mathrm{F}}+\sqrt{p_{\mathrm{F}}^{2}+m_{\mathrm{eff}}^{2}}}{m_{\mathrm{eff}}}
    \right)
  \right].
\end{multline}
However, $p_{\mathrm{F}}$ is also the function of $r$ in Eq.~(\ref{eq:diffeqphi}).
This fact can be accounted for by using the condition for the chemical
equlibrium in a spatially inhomogeneous system (see, e.g., Ref.~\cite{LanLif69}),
\begin{equation}\label{eq:chemequil}
  \mu=\mathrm{const},
  \quad
  \mu=\sqrt{p_{\mathrm{F}}^{2}+m_{\mathrm{eff}}^{2}}.
\end{equation}
Based on Eq.~(\ref{eq:chemequil}), we fix the chemical potential
in the center of the neutrino cluster $\mu_{0}=\sqrt{p_{\mathrm{F}}^{2}(r=0)+m_{\mathrm{eff}}^{2}(r=0)}$.

Using the dimensionless variables,
\begin{equation}
  z=Mr,
  \quad
  \Phi=\frac{zg\left\langle \phi\right\rangle }{m},
\end{equation}
as well as Eq.~(\ref{eq:chemequil}), we rewrite Eq.~(\ref{eq:diffeqphi})
in the form,
\begin{align}\label{eq:diffeqPhi}
  \frac{\mathrm{d}^{2}\Phi}{\mathrm{d}z^{2}}= & \Phi-\frac{(z-\Phi)}{2\pi^{2}\chi^{2}z^{2}}
  \nonumber
  \\
  & \times
  \left[
    z\mathcal{M}\sqrt{z^{2}\mathcal{M}^{2}-(z-\Phi)^{2}}-(z-\Phi)^{2}\ln
    \left(
      \frac{\sqrt{z^{2}\mathcal{M}^{2}-(z-\Phi)^{2}}+z\mathcal{M}}{z-\Phi}
    \right)
  \right].
\end{align}
where $\chi=\tfrac{M}{gm}$ and $\mathcal{M}=\tfrac{\mu_{0}}{m}$.

We suppose that both $\left.\left\langle \phi\right\rangle \right|_{r=0}$
and $\left.\tfrac{\mathrm{d}}{\mathrm{d}r}\left\langle \phi\right\rangle \right|_{r=0}$
are finite. Thus, Eq.~(\ref{eq:diffeqPhi}) should be supplied with
the initial condition: $\left.\Phi\right|_{z=0}=0$ and $\left.\tfrac{\mathrm{d}}{\mathrm{d}z}\Phi\right|_{z=0}=\tfrac{g}{m}\left.\left\langle \phi\right\rangle \right|_{r=0}$.
It is more convenient to use $\mathcal{M}$ and $\Pi_{0}=p_{\mathrm{F}}(r=0)/m$
as the alternative initial condition. In this case, $\left.\tfrac{\mathrm{d}}{\mathrm{d}z}\Phi\right|_{z=0}=1-\sqrt{\mathcal{M}^{2}-\Pi_{0}^{2}}$.
Of course, one requires that $\mathcal{M}>\Pi_{0}$.

Equation~(\ref{eq:diffeqPhi}) is integrated numerically from $z=0$ to the
point when
\begin{equation}\label{eq:pFr}
  \Pi(r)=\frac{p_{\mathrm{F}}(r)}{m}=\sqrt{\mathcal{M}^{2}-
  \left(
    1-\frac{\Phi}{z}
  \right)^{2}},
\end{equation}
drops to zero. This point means the radius of a neutrino cluster.

\section{Neutrino cluster cooling\label{sec:COOL}}

In this appendix, we suggest a possible channel for the cooling a neutrino cluster down.

If we assume that a neutrino cluster appears in the early Universe after the electroweak phase transition at $T \lesssim 100\,\text{GeV}$, neutrinos can interact with both the scalar field $\phi$ and other particles within the standard model. Such interactions can provide the cooling of the neutrino gas inside the cluster to the temperature of the outer medium.

We suppose that the cooling is owing to the \v{C}erenkov radiation of plasmons by primordial neutrinos~\cite{Raf96}, $\nu\to\nu\gamma$. This process is allowed in the standard model even for massless neutrinos. Other possibilities include the \v{C}erenkov radiation of millicharged neutrinos and neutrinos possessing nonzero magnetic moments~\cite{Raf96}. We do not exclude these cases, however they are not considered here.

We take the maximal density of a neutrino cluster in the present universe $n_\text{clust}^{(\mathrm{now})} = 10^9\,\text{cm}^{-3}$ derived in Ref.~\cite{SmiXu22} (see also Fig.~\ref{fig:f1a}). If we consider only electron neutrinos forming the cluster, their density in C$\nu$B is $n^{(\mathrm{now})} = 56\,\text{cm}^{-3}$ nowadays~\cite[p.~135]{GorRub11}. The corresponding densities in the early universe are $n_\text{clust} = n_\text{clust}^{(\mathrm{now})}/a^3$ and $n = n^{(\mathrm{now})}/a^3$, where $a$ is the scale factor of the Friedmann--Robertson--Walker metric.

We suppose that the contraction of the neutrino gas in a cluster is adiabatic. Despite it is the overestimation since some energy transmission through the cluster border is possible, we admit it assuming that the formation of a cluster is fast. The final neutrino temperature in a cluster is $T_\mathrm{clust} = T \left(\tfrac{n_\text{clust}}{n}\right)^{\gamma-1} = 6.7\times10^4 T$, where $T$ is the temperature of primodrial plasma at the time of the cluster formation and $\gamma = 5/3$ is the heat capacity ratio of the neutrino gas.

Based on the results of Ref.~\cite{DolNiePal96}, the energy loss in the plasmon emission is estimated as 
\begin{equation}
  \dot{E} \sim \frac{G_\mathrm{F}^2\omega_p^6}{64\pi^2\alpha_\mathrm{em}},
\end{equation}
where $G_\mathrm{F} = 1.17\times 10^{-5}\,\text{GeV}^{-2}$ is the Fermi constant,  $\omega_p \sim \alpha_\mathrm{em}^{1/2} T$ is the plasma frequency in a hot relativistic matter~\cite{BraSeg93,DvoSem14}, and $\alpha_\mathrm{em} = 7.3\times 10^{-3}$ is the fine structure constant. Here, we assume that the emission of plasmons happens mainly from the cluster surface to get the strongest lower bound on the plasma temperature.

The condition of the successful cluster cooling down to the temperature of the outer primordial plasma is 
\begin{equation}\label{eq:condcool}
  \frac{T_\mathrm{clust}}{\dot{E}} < H^{-1},
\end{equation}
where $H = T^2/M_\mathrm{Pl}^*$ is the Hubble parameter, $M_\mathrm{Pl}^* = M_\mathrm{Pl}/1.66\sqrt{g_*}$, $M_\mathrm{Pl} = 1.2\times 10^{19}\,\text{GeV}$ is the Planck mass, and $g_*$ is the effective number of relativistic degrees of freedom. We take that $g_* \approx 80$ at $T \lesssim 100\,\text{GeV}$~\cite[p.~409]{GorRub11}.

Based on the above estimates, Eq.~\eqref{eq:condcool} is rewritten the form, $T>20\,\text{GeV}$. Thus, a neutrino cluster which appears in the epoch when $20\,\text{GeV} < T < 100\,\text{GeV}$, cools down owing to the \v{C}erenkov radiation of plasmons soon after its formation. The temperature of the neutrino gas in such a cluster then reaches $T_{\text{C}\nu\text{B}}=1.95\,\text{K}$ in the present times owing to the universe expansion.

Note that, if we increase the maximal number density in a cluster, the temperature interval in the early universe, necessary for the cluster cooling, is very narrow. It means that, the conditions required for the superfluidity appearance are not fulfilled in clusters with rather high density of neutrinos unless other more efficient cooling channels are proposed. 

\section*{Acknowledgments}


I am thankful to V.~B.~Semikoz for useful discussions.

\end{document}